\documentclass{article}
\usepackage{spconf,amsmath,graphicx}
\usepackage[hidelinks]{hyperref}
\usepackage{booktabs}
\usepackage{amssymb}

\title{Personalized Task Load Prediction in Speech Communication}

\name{
    Robert P. Spang$^{1}$,
    Karl El Hajal$^{2,3}$,
    Sebastian Möller$^{1,4}$,
    Milos Cernak$^{3}$
    \thanks{This work was financially supported by the German Federal Ministry of Education and Research through Software Campus grant 01IS17052 (KOVOSEQ)}
}
\address{
    $^{1}$Quality and Usability Lab, Technische Universität Berlin, Germany,\\
    $^{2}$\'Ecole Polytechnique F\'ed\'erale de Lausanne (EPFL), Lausanne, Switzerland,\\
    $^{3}$Logitech Europe, Lausanne, Switzerland,\\
    $^{4}$German Research Center for Artificial Intelligence (DFKI), Berlin, Germany
}
\begin{document}

\maketitle

\begin{abstract}
    Estimating the quality of remote speech communication is a complex task influenced by the speaker, transmission channel, and listener. For example, the degradation of transmission quality can increase listeners' cognitive load, which can influence the overall perceived quality of the conversation. This paper presents a framework that isolates quality-dependent changes and controls most outside influencing factors like personal preference in a simulated conversational environment. The performed statistical analysis finds significant relationships between stimulus quality and the listener's valence and personality (agreeableness and openness) and, similarly, between the perceived task load during the listening task and the listener's personality and frustration intolerance. The machine learning model of the task load prediction improves the correlation coefficients from 0.48 to 0.76 when listeners' individuality is considered. The proposed evaluation framework and results pave the way for personalized audio quality assessment that includes speakers' and listeners' individuality beyond conventional channel modeling.
\end{abstract}


\section{Introduction}
\label{sec:intro}

    Task load is a measurement of human performance that broadly refers to the difficulty levels an individual encounters when executing a task~\cite{Zimmerman2011}. With the recent shift to hybrid work, virtual interaction can come with a cognitive cost for creative idea generation~\cite{brucks2022virtual}. However, cognitive load prediction is still an unsolved research (and engineering) problem, although recent audio analysis~\cite{elbanna22_interspeech} and video quality assessment~\cite{Qi2022,Shaoguo22,Ao-Xiang22} findings are promising. 
    
    When investigating the perception of real-time conversations, such as quality assessments, researchers usually end up with two options: asking people how they liked it or how they felt about it after the fact (following, for example, ITU-T P.800~\cite{p800} in case of speech quality assessment). Such subjective listening tests give us an individual rating, but at the cost of temporal resolution, and it is only an overall score for the entire trial. Alternatively, we can employ objective intrusive or non-intrusive methods pre-trained on thousands of ratings. Such systems allow us to evaluate a comm link in real-time, suggesting reasons for audio distortions~\cite{elhajal22_interspeech} but at the cost of individuality. Since those models are trained on average ratings for many individuals, their output is a mean-opinion-score that might be true for a grand average but neglects individual differences in users.
    
    \begin{figure}
        \centering
        \includegraphics[width=1\columnwidth]{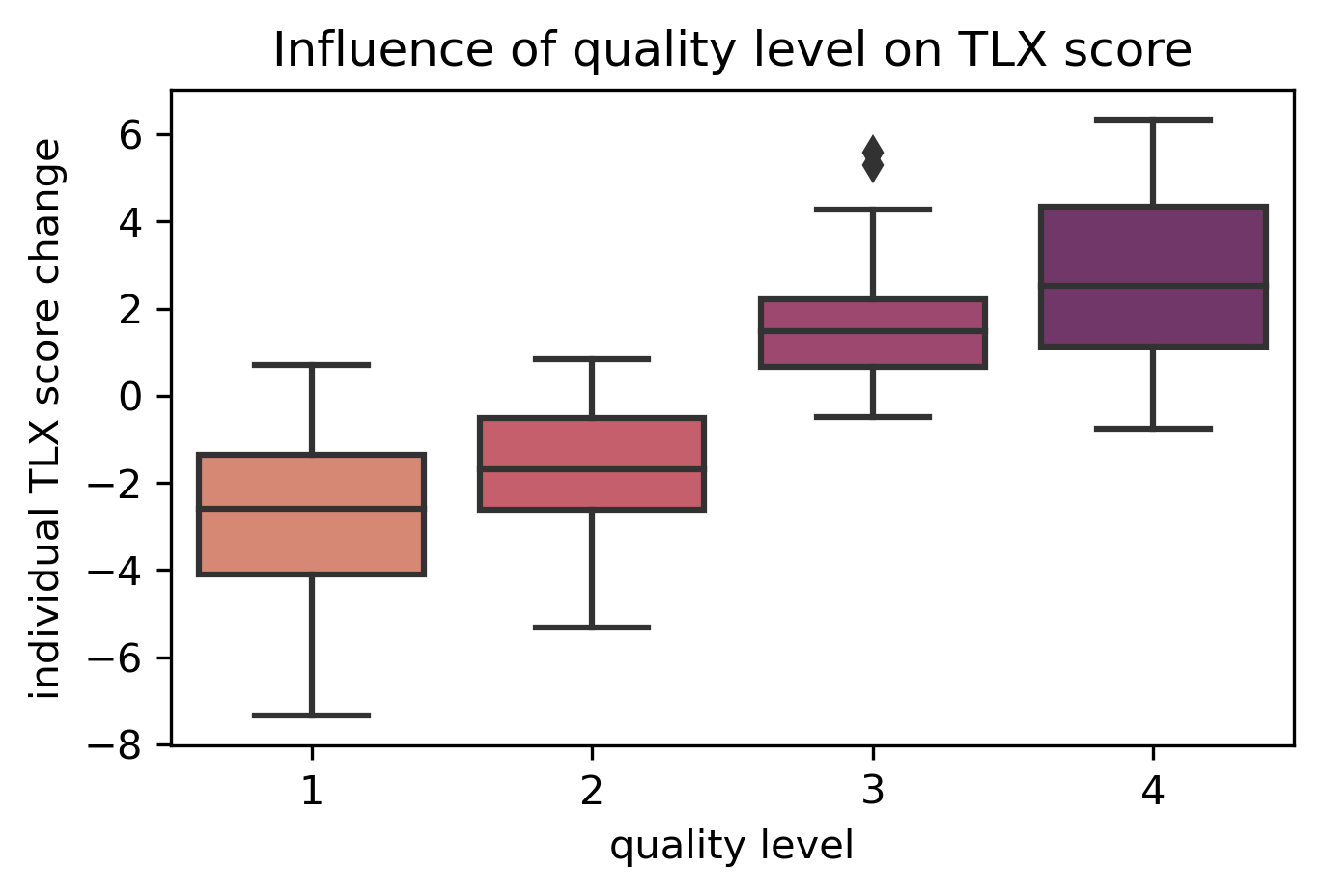}
        \caption{Relationship between audio quality and perceived task load (TLX), normalized per user. The lesser the quality, the higher the users' mental load.}
        \label{fig:tlx_quality_relation}
    \end{figure}
    
    We hypothesize that overall quality is factorized by the speaker (spoken content, speaker's familiarity), channel (coding distortions, transmission drops), and listener (personality, mood). For example, the degradation of transmission quality can increase listeners' cognitive load, which can influence the overall quality of the conversation. Figure~\ref{fig:tlx_quality_relation} shows a direct influence of a stimulus' quality on the listener's perceived task load. This relationship was recorded in the user study described in~\ref{ssec:userstudy}. As the stimuli' quality drops (level 1 was best and 4 was worst), listeners perceive a greater mental load than the stimulis' best quality. More mental resources are needed to be able to understand the content of the stimuli. Task load is directly related to quality and other stimuli-dependent measures of the users' perception.
     
    In line with other reports, the audio quality -- task load relationship allows us to use knowledge regarding the stimuli's quality to estimate users' perceived mental load while, e.g., following a conversation~\cite{van2014classification,govender2018,scherer2002acoustic,huttunen2011effect,schuller2014interspeech}. In our example above, this simple relation explains a good amount of variance ($R^2: 0.235$). However, we want to investigate if additional information about the individuals can increase the amount of variance explained. For example, information on the users' personality and frustration (in)tolerance might help to estimate better someone's perceived task load.
    
    In this work, we thus explore cognitive task load prediction and claim that its personalization, i.e., incorporating the listener's personality, frustration, and emotional levels, can significantly impact its accuracy. We validate our personalized task load prediction by solid statistical analysis and justify further the findings by machine learning experiments. Intrusive audio quality assessment methods are perceptually limited~\cite{manocha22_interspeech}, and we thus base our experiments on non-intrusive speech quality assessment methods.
    
    We hypothesize that knowledge of the listeners' personalities improves the prediction of their perception. This manuscript demonstrates an improvement of a subjectively perceived dimension throughout a simulated video telephony scenario. The simulation allows us to isolate quality-dependent changes and control most outside influencing factors like personal preference in a real conversational environment. We showcase good prediction performance using the (degraded) audio channel alone as an input and then gradually improve this prediction by adding more personality/trait dimensions of the respective listener. While focusing on perceived task load, we discuss applications to various dimensions, like perceived quality, valence, or arousal.

\section{Methods}
\label{sec:evaluationmethod}

    To increase the prediction accuracy of any perception dimension, we need a theoretical model, a valid approach, a prediction system, and data to test. While we outlined the theoretical model in the introduction, the following describes how we obtained data for this experiment, how we tested the concept statistically, and we describe the machine learning pipeline that we employ to predict subjectively perceived task load.
    
\subsection{Data collection}
\label{ssec:userstudy}

    We collected data through a controlled lab study at the Technical University of Berlin, including 46 participants (30 years of median age, 22-57y), ethics board of Faculty IV Electrical Engineering and Computer Science approval ID: \textit{FR\_2022\_03}).
    Participants were handed the NEO-FFI questionnaire~\cite{borkenau1993neo}, a standardized inventory to assess the personality dimensions agreeableness, openness, conscientiousness, extraversion, and neuroticism.
    Additionally, we employed the Frustration Discomfort Scale (FDS)~\cite{harrington2005frustration} to assess one's ability to accept and cope with frustration. Both scales were used to quantify the participants' personalities.
    
    In the main task, participants watched a series of short video clips (10-12s each).
    The clips show realistic quality distortions~\cite{storytimedata2022}; quality levels were chosen at random. Participants watched thirty video clips and answered questionnaires after each trial: Task-load ratings were obtained using the NASA-TLX questionnaire~\cite{hart1988development}, and valence ratings were obtained through the affect grid. Participants were compensated for their time with 15 Euros/h.
    This procedure yielded thirty task-load scores per participant. However, due to one person not responding to the personality questionnaires and several missing values (treated as missing completely at random), we obtained 1,279 complete samples from 45 participants.
    
    \subsection{Statistical analysis}
    \label{ssec:methodsstatsanalysis}
            
    The dataset contains TLX score values ranging $[0, 20]$, valence ratings ranging $[0, 1]$, audio-MOS scores (obtained using POLQA~\cite{beerends2013perceptual}) ranging $[0, 4.79]$, frustration intolerance score ranging $[1, 5]$, and five personality scores, ranging $[1, 5]$. All values are continuous.
    A path analysis, a Structural Equation Model without a measurement model, used to describe directed dependencies among a set of variables, allows us to account for possible mediation effects between valence and the task load.
    
\begin{figure}
        \centering
        \includegraphics[width=1\columnwidth]{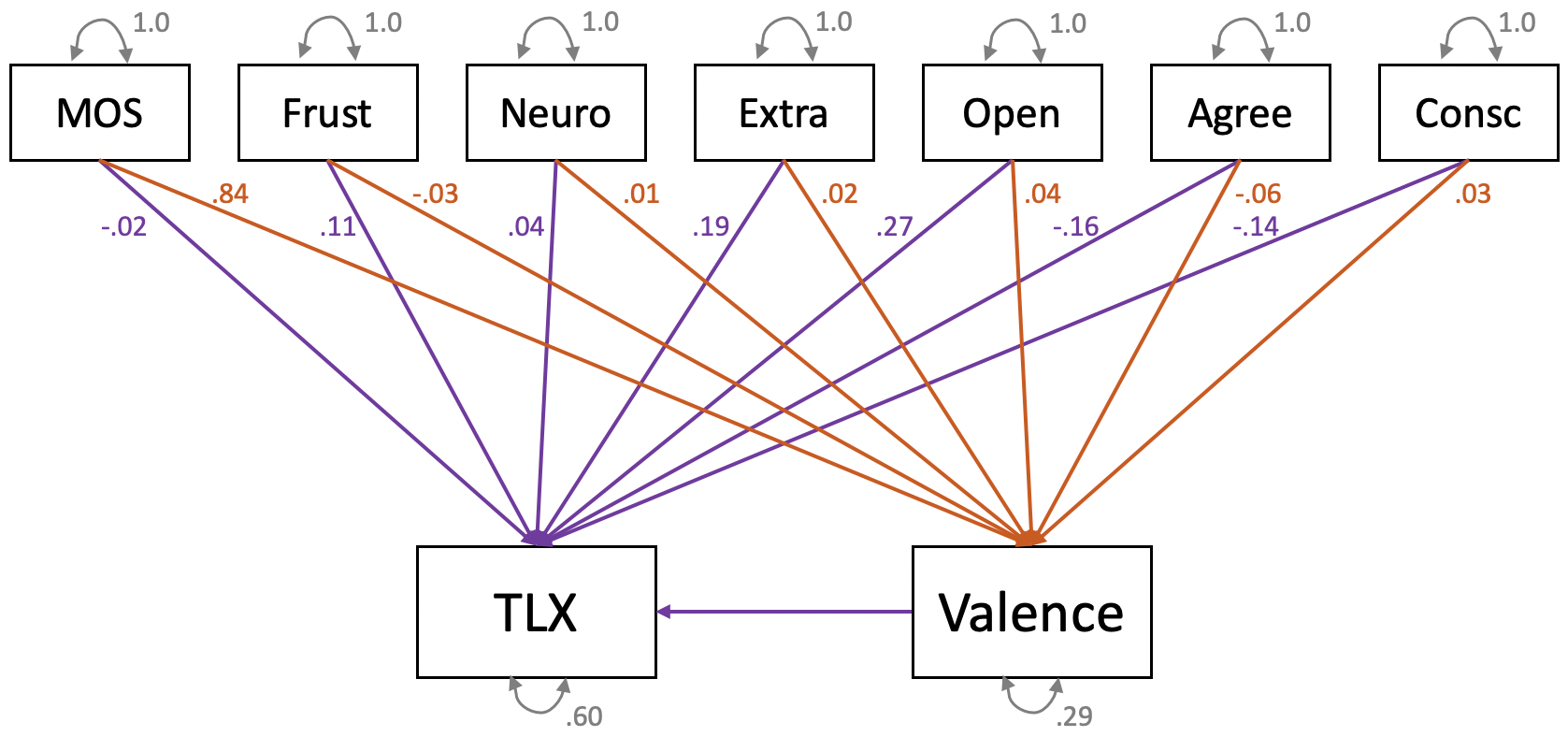}
        \caption{Structural equation Model, path diagram coefficients are standardized $\beta$-values.}
        \label{fig:sem_model_graph}
    \end{figure}

    We ran a path analysis to understand the relationship between stimulus quality (audio MOS), perceived task load (TLX score) and valence, and individual characteristics of our participants (frustration intolerance and all personality dimensions).
    The analysis was carried out using lavaan~\cite{rosseel2012lavaan} with the Robust Maximum Likelihood estimator in Jamovi/R~\cite{jamovi2020jamovi,team2013r,gallucci2021semlj}, see Figure~\ref{fig:sem_model_graph}.
            
\subsection{Machine learning-based analysis}
\label{ssec:mlmethod}

    The machine learning experiment assesses whether providing a model with information about each individual would help predict the task load participants perceived for a certain stimulus.
    Therefore, we trained models to predict TLX scores using a raw audio signal and information about the listener as input. The latter consists of frustration intolerance, personality measures, and participants' perceived valence ratings. We experiment with all permutations of these dimensions to see which would yield the most explanatory power in predicting TLX scores.
    
\begin{figure}
        \centering
        \includegraphics[width = 0.7\columnwidth]{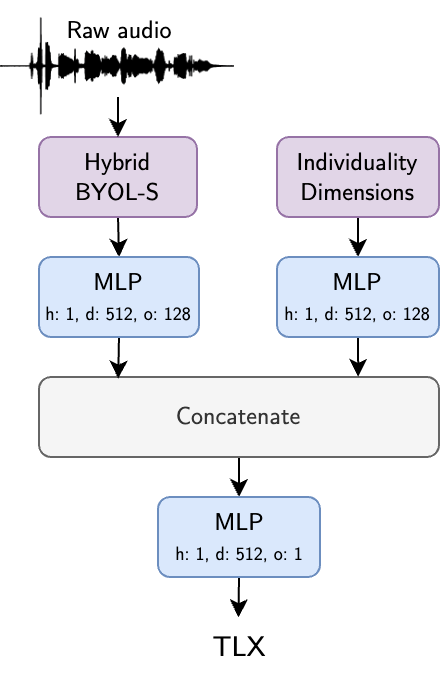}
        \caption{ML model architecture. h: number of hidden layers, d: the dimension of each hidden layer, o: the size of each output layer. Individuality dimensions: five personality dimensions + frustration intolerance score.}
        \label{fig:ml_model}
    \end{figure}

    The model used to perform these experiments is visualized in Figure~\ref{fig:ml_model}.
    First, features are extracted from the raw audio using the Hybrid BYOL-S/CvT model from \cite{elbanna2022byol}, which was shown to be effective for a wide range of speech tasks. We extract utterance-level embeddings, meaning that the whole raw audio is reduced to a 1D feature vector of dimension 2048. The Hybrid BYOL-S features and the chosen individuality dimensions then go through Multilayer Perceptrons (MLP), each of which has one hidden layer of dimension 512 and an output layer of size 128. The ReLU layer is used as a non-linearity. The two outputs are concatenated and fed to a final MLP that has one hidden layer of dimension 512 and outputs the predicted TLX values.
    
    The data is normalized for these experiments. The model is trained using the ADAM \cite{kingma2014adam} optimizer, a batch size of 8, a learning rate  $\alpha = 1\mathrm{e}{-5}$, and the Mean Absolute Error (MAE) loss.
    We calculate the Pearson Correlation Coefficient (PCC) and MAE metrics to assess the model's performance. We address potential overfitting by using a 5-fold cross-validation for evaluation and calculate the mean PCC and MAE values across all folds.
    
\section{Results}
\label{sec:results}
\label{ssec:statsanalysis}
    
    The structural model estimates two variables; valence as a mediator and TLX as an endogenous variable. For valence, we modeled frustration intolerance, all five personality dimensions, and the audio MOS as exogenous variables. To estimate TLX, we used the same exogenous variables as for valence but also added valence as an additional input variable. We further specified several covariances between the exogenous variables to optimize the overall model fit. See Section~\ref{ref:data_availability} for full covariance and correlation tables.
    
    The resulting structural model obtained good fit with $\chi^2$ = 85.049; df = 11; p $<0.001$; SRMR = 0.054; CFI = 0.977; RMSEA = 0.073. Given our dataset, these fit indices attest general validity of the modeled structure. Therefore, it is a valid tool for investigating our hypothesis. The variance explained is $R^2=0.71$ for valence and $R^2=0.404$ for TLX. Table~\ref{tab:sem_analysis_estimates} provides resulting parameter estimates.
    
    For valence, three of the seven exogenous variables significantly influence the model: the MOS score of each clip has the strongest influence with a $\beta$ of 0.839. Regarding the perceived task load, all but two variables contribute significantly to the resulting regression model. The personality dimension neuroticism only tends to influence the TLX, while the audio quality does not contribute significantly to the TLX. The mediator valence has the strongest $\beta$ loading of -0.551.
    
    \begin{table}[ht!]
    \caption{Parameter estimates of the structural model. Abbreviations are val: valence, mos: MOS-score of the audio clip, agree: agreeableness, open: openness, frust: frustration intolerance score, consc: conscientiousness, extra: extraversion, neuro: neuroticism, and TLX: task load index.}
    \label{tab:sem_analysis_estimates}
    \centering
    \begin{tabular}{lcrcccc}
        \toprule
        &             &          & Est  & SE & p      & $\beta$ \\

        \midrule
        val & $\thicksim$ & mos      & 0.156 & 0.003 & $\mathbf{<0.001}$ & 0.839 \\
        val & $\thicksim$ & agree    & -0.064 & 0.026 & \textbf{0.012}   & -0.058 \\
        val & $\thicksim$ & open     & 0.043 & 0.018 & \textbf{0.014}    & 0.044 \\
        val & $\thicksim$ & frust    & -0.022 & 0.033 & 0.515   & -0.033 \\
        val & $\thicksim$ & consc    & 0.029 & 0.022 & 0.185    & 0.026 \\
        val & $\thicksim$ & extra    & 0.023 & 0.018 & 0.204    & 0.021 \\
        val & $\thicksim$ & neuro    & 0.008 & 0.015 & 0.570    & 0.011 \\
        
        \midrule
        TLX     & $\thicksim$ & val     & -8.334 & 0.673 & $\mathbf{<0.001}$ & -0.551 \\
        TLX     & $\thicksim$ & open    & 3.962 & 0.371 & $\mathbf{<0.001}$  & 0.269 \\
        TLX     & $\thicksim$ & extra   & 3.120 & 0.378 & $\mathbf{<0.001}$  & 0.189 \\
        TLX     & $\thicksim$ & agree   & -2.629 & 0.420 & $\mathbf{<0.001}$ & -0.156 \\
        TLX     & $\thicksim$ & consc   & -2.427 & 0.475 & $\mathbf{<0.001}$ & -0.144 \\
        TLX     & $\thicksim$ & frust   & 1.104 & 0.310 & $\mathbf{<0.001}$  & 0.110 \\
        TLX     & $\thicksim$ & neuro   & 0.489 & 0.278 & 0.079     & 0.042 \\
        TLX     & $\thicksim$ & mos     & -0.056 & 0.124 & 0.652    & -0.020 \\
        \bottomrule
    \end{tabular}
    \vspace{-9pt}
\end{table}
    
    \label{ssec:mlanalysis}
    \begin{table}[ht!]
    \caption{Results showing the model's ability to predict the task load index using different individuality dimensions as input. The metrics used are mean PCC ($\uparrow$) and MAE ($\downarrow$) across all folds.}
    \vspace{4.5pt}
    \label{tab:machine_learning_results}
    \centering
    \begin{tabular}{lccc}
        \toprule
        Input &  PCC & MAE \\
        \midrule
        Audio & 0.48 & 0.73 \\
        \midrule
        Audio + Valence & 0.58 & 0.66 \\
        Audio + Personality & 0.68 & \textbf{0.59} \\
        Audio + FDS & \textbf{0.69} & \textbf{0.59} \\
        \midrule
        Audio + FDS + Valence & 0.69 & 0.59 \\
        Audio + Personality + Valence & 0.70 & 0.57 \\
        Audio + FDS + Personality & \textbf{0.74} & \textbf{0.53} \\
        \midrule
        Audio + FDS + Personality + Valence & \textbf{0.76} & \textbf{0.51} \\
        \bottomrule
    \end{tabular}
    \vspace{-9pt}
\end{table}
    
    The machine learning-based analysis results are shown in Table~\ref{tab:machine_learning_results}. Feeding individuality information into the model as input helps it to predict perceived TLX significantly better than the audio-only experiment. When only one set of individuality dimensions is fed to the model, we can see that FDS values and personality dimensions are comparably good at improving the model's results. Meanwhile, adding the valence dimension also shows a significant improvement. When two individuality dimensions are used as input, combining the FDS and personality scores allows the model to improve further. Finally, combining all three dimensions allows the model to achieve its overall best performance.

\section{Discussion}
\label{sec:discussion}
    
    The raw audio quality alone determines the perceived valence of the users to a large degree. Only the personality dimensions of agreeableness and openness significantly influence the valence ratings further. However, comparing the $\beta$-loadings of the two personality dimensions with the MOS score reveals a rather small effect of personality on valence. The other three dimensions and the frustration discomfort score do not provide statistically meaningful explanations for the valence dimension. Besides negligible other influences, valence is mostly determined through the content's quality. In comparison, a distorted connection decreases the perceived positivity of the listeners, and good quality yields significantly more positive perceptions among participants.
        
    Regarding TLX, the more valence people perceive, the lesser the mental load they report. In contrast, a low valence score is correlated with an increased mental load. MOS score does not contribute significantly to the task load estimation, although it is the strongest factor in a model without valence as a mediator. However, most of its explanatory power seems to be masked by valence. Concluding the statistical analysis, the information about personality and frustration intolerance help explain much variance in the listeners' responses. Adding information about someone's traits provides a much clearer understanding of their responses and how they felt during the listening task. During the modeling phase, we also tested a variant using TLX as a mediator to valence. However, such a model showed less favorable fit indices, so we settled with the model described above.
        
    Similar to our statistical analysis, the analysis of our ML-model approach shows strong gains by adding personality and frustration intolerance dimensions over only using the audio files alone to predict the perceived task load. Both the correlation of the model's output got stronger, while the average error shrank. Adding knowledge of the participant's ratings of the clips enables us to generate individualized predictions. In applications, these individualized signal analyses allow for targeted estimations instead of the same output offered to everyone if only the input quality is the same.

\subsection{Conclusions}
\label{ssec:conclusion}
        
    We have provided a framework to demonstrate how individuality metrics, here personality and frustration intolerance, greatly improve signal analyses of perception dimensions. We conducted a statistical and ML-model-based analysis and found several significant influences, on the one hand, increased correlation coefficients (from 0.48 to 0.74) and reduced errors (from 0.73 to 0.53 score points) in the prediction of perceived task load on the other hand. While the dataset and presented analyses are a good case study to demonstrate the relevance of subject knowledge, this manuscript is foremost a motivation towards individualized perception estimation.
            
    Our analyses are suitable for discussing individuality in signal processing and perception predictions. However, more studies must be conducted to substantiate our claim and identify the areas of strongest influence.
    Further investigations could use crowd-sourced research to obtain a magnitude more responses than this first introduction study has. Because the present dataset is fairly small, overfitting is a potential limitation of this work. However, we will continue investigating the stated relationships and substantiate our findings over time.

\section{Data Availability \& Acknowledgment}
    \label{ref:data_availability}
    The dataset, the statistical analysis files, and the ML pipeline described are publicly available\footnote{\href{https://osf.io/rxbpv/}{https://osf.io/rxbpv/}}.

    We thank Kerstin Pieper for her support with recording the videos and, together with Martin Burghart and Leon Schreiber, for their support in conducting the subjective evaluation study.

\bibliographystyle{IEEEbib}
\bibliography{references}

\end{document}